\documentclass[twocolumn,showpacs,preprintnumbers,amsmath,amssymb]{revtex4}
  \usepackage{graphicx}
  \usepackage{dcolumn}
  \usepackage{bm}
 
\begin{document}

\title{The role of thermal disorder for magnetism and the $\alpha - \gamma$ transition in Cerium; 
Results from density-functional theory.}

  \author{T. Jarlborg}

  \affiliation{
  DPMC, University of Geneva, 24 Quai Ernest-Ansermet, CH-1211 Geneva 4,
  Switzerland
  \\
  }
 
 
\begin{abstract}

 The electronic structures of fcc Ce are
 calculated for large supercells with varying disorder by use of density-functional theory. Thermal
 disorder induces fluctuations of the amplitude of
 the magnetic moments and an increase the average moments in the high-volume
 phase.  The ferro-magnetic solutions move towards lower volume than in calculations for
 the perfectly ordered lattice. Therefore, disorder contributes via entropy to the stabilization of the
 $\gamma$ phase at high $T$, and it is important for an understanding of the $\alpha-\gamma$
 transition. Core level spectroscopy would be a mean to detect disorder through
 the spread of Madelung shifts and local exchange splittings.
  \end{abstract}
 
 
  \maketitle

\section{Introduction}

The isostructural $\alpha-\gamma$ transition in fcc Ce attracts a renewed interest 
\cite{lip,lipp,casa,bied,lana,tran,decr,jeong}.
Ce undergoes a volume reduction of up to 17 percent from the magnetically disordered $\gamma$-phase to
a non magnetic (NM) low volume $\alpha$-phase at low pressure, $P$, at a temperature $T$ of about 300 K, 
even though a competing dhcp phase, $\beta$,
appears \cite{fran,pic} at lower $T$ for zero $P$. The transition moves to higher $T$ for higher $P$; at $\sim$ 20 kbar
it reaches almost 600 K with a vanishing volume reduction. 
Several models have been proposed to drive the transition, the Kondo volume-collapse model \cite{alle},
a Mott transition \cite{joha}, correlation and entropy \cite{bied,lana,amad}, or via standard density-functional theory (DFT)
bands with entropies \cite{ce,wang}. The T-dependence is the unusual feature of the transition.
At $T=0$ only the $\alpha$-phase is stable. Nevertheless, a recent work proposed that the DF-potential
should be replaced by a different hybrid exchange-correlation potential, with enhanced correlation, 
since calculations at $T=0$ compare favorably with measured extrapolation of to a negative transition pressure \cite{casa}.
There are two groups of models containing more or less of correlation. 
Either they conclude that Ce is a strongly correlated
system, where DFT is not sufficient, or otherwise they suggest that DFT is applicable, but that the
band structure information has to be complemented by entropy \cite{ebj,wany}.  
In fact, DFT-calculations based on the local spin-density
approximation (LSDA) \cite{lsda}, and  the generalized gradient approximation (GGA) \cite{gga},
both give a qualitatively correct account of the T-dependence of the $\alpha-\gamma$
transition if the entropies are included \cite{ce,wang}, but GGA
is quantitatively better 
\cite{ce}.
None of these DFT-potentials include particular on-site correlation beyond the
normal correlation within the electron gas. 

Independently of this, it has been shown that effects from thermal
disorder and zero-point motion (ZPM) are important for the band structure and properties in materials
with sharp density-of-states (DOS) variations near the Fermi energy ($E_F$). 
In such cases it is necessary to include disorder into the electronic structure calculations for 
a correct description of the physical properties or spectroscopic responses \cite{fesi,fege,dela,bron}. 
Magnetic manifestations of coupling between lattice distortions and electronic structures show up 
as spin-phonon coupling in cuprates \cite{tj7}
and spin-lattice interactions giving invariant thermal expansion in INVAR materials \cite{inva}.
Ce is, a priori, a system where disorder could be important, because the DOS of the f-band raises sharply at $E_F$, 
and the lattice is fairly soft, which makes large distortion amplitudes
of atomic vibrations already at low T. 
  In the present work we investigate the importance of lattice disorder for the properties of Ce and its
transition between the magnetic and the non-magnetic phase. 

\section{Results and discussion}

\subsection{Lattice disorder and magnetism}
 
Self-consistent linear muffin-tin orbital (LMTO) band calculations \cite{lmto} have been
made for 32-atom unit cells, 2x2x2 extensions of the cubic fcc 4-atom cell,
 of fcc Ce for several lattice constants ($a_0$) between 4.85 and 5.42 ~\AA. 
The GGA potential is used \cite{gga}. Calculations
are also made for the ordinary (ordered) fcc 1 atom/cell for $a_0$ between 4.75 and 5.4 ~\AA. 
The number of k-points is 75 in half of the Brillouin zone (BZ) for the large cell, and 505 in 1/48 BZ for the
small cell.  The 6s, 5p, 5d and 4f
valence electrons are included in the basis. 
Magnetic moments and DOS functions are very similar in these two sets
of calculations for ordered structure.

 Each atomic
position in the 32 atom/cell is randomly displaced in the calculations with thermal disorder so that the averaged
displacement amplitudes follow a Gaussian distribution function with width $\sigma$ \cite{fesi}. This distribution
is valid at not too large $T$, when there are no correlated movements of the neighboring atoms \cite{grim}.
The average lattice displacement $u$ is related to $T$ as
\begin{equation}
u^2 = 3 k_B T/ \mathcal{M} \omega^2
\label{equT}
\end{equation}
where $\mathcal{M}$ is the atomic weight and $\omega$ an averaged phonon frequency.
ZPM remains at $T$ well below the 
Debye temperature, $\Theta_D$, and 
$u^2$ is never smaller than $3 \hbar \omega / M \omega^2$ \cite{zim,grim}.
From the measured $\Theta_D$ of about 115 K for $\gamma$-Ce and 160 K for the high-$P$ room temperature (RT)
$\alpha$-phase \cite{lip} we can estimate that $\sigma_{ZPM} = u_{ZPM}/a_0$ is of the order
0.01-0.013, and that $\sigma_T$ at RT is about 0.021-0.028 for $\gamma$ and $\alpha$, respectively.
The band calculations are made for several
disordered 32-cell configurations with $\sigma$ from 0.021 up to 0.063, which correspond
to a temperature range between approximately 200 K and 800 K. For comparison we note that $u$
would be about 0.22 of the atomic sphere radius according to the Lindemann criterion for the melting temperature, $T_m$
\cite{grim},
i.e. $\sigma$ would be of the order 0.086. This fits with our $T$-calibration of $\sigma$,
since $T_m$ is near 1050 K for Ce \cite{fran}.

The band structure is sensitive to disorder (and ZPM) because of the fluctuations of the 
potential in a vibrating disordered lattice. 
The Coulomb potential $v_i(r)$ at a point $r$ within a site $i$ is
\begin{equation}
v_i(r) = - \sum_j Z_j/|r-R_j| + \int_0^{\infty} \rho(r')/|r-r'| d^3r'
\end{equation}
where $Z_j$ are the nuclear charges on sites $j$, $\rho(r)$ is the electron charge density,
and the sum and integral cover all space.
The contribution to $v_i(r)$ from its own site (with radius $S_i > r'$) can be separated from the
contribution from the surrounding lattice through the technique of Ewald lattice summation \cite{zim}:
\begin{equation} 
v_i(r) \approx -Z_i/r + \int_0^{S_i} \rho(r')/|r-r'|d^3r' + M_i
\label{eqV}
\end{equation}
Thus, the Coulomb interaction with the outside lattice is condensed into a Madelung shift, $M_i$. This shift is
 identical for all sites if the lattice is perfectly ordered. But different sites have different $M_i$
in a disordered lattice, partly because of the local differences in atomic positions and partly
because of the charge transfers induced by the disorder. Thus,
the potentials at different sites are slightly different and they
vary in time. Phonons are very slow compared to the electronic time scale and the electronic
structure can relax adiabatically.
The band results for two different configurations with almost the same $\sigma$
are comparable, which indicates that the 32-atom cells are large enough for simulation of disorder. 
Other details of the calculational method can be found in refs.
\cite{lmto,ce}.

\begin{figure}[h]
  \begin{center}
  \includegraphics[width=8cm,height=6cm]{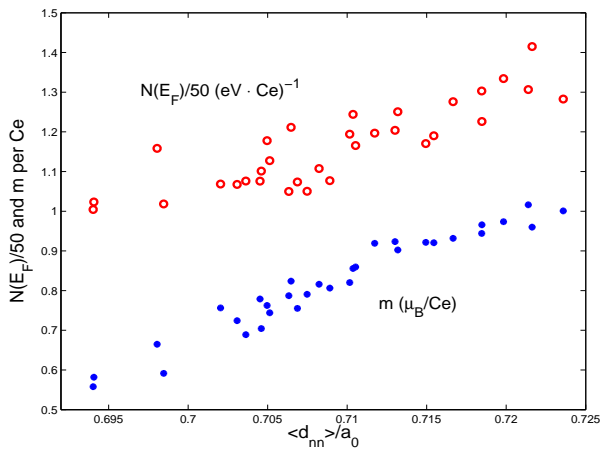}
  \end{center}
  \caption{Correlation between the averaged nearest-neigbor distances, $<d_{nn}>$ (in units of the lattice constant), the
  DOS at $E_F$ (red open circles, in states per $eV \cdot Ce$/50), 
  and local magnetic moments (blue points, in $\mu_B/Ce$), calculated for a 32-site cell with a
  disorder of $\sigma$=0.04 at the lattice
  constant $a_0 =$ 5.29 \AA. "Compressed" sites with small $<d_{nn}>$ have small
  $N(E_F)$ and $m$.
  Oppositely, for "isolated" sites with large $<d_{nn}>$ the $N(E_F)$ and m are the highest. } 
  \label{fig1}
  \end{figure}

\begin{figure}[h]
  \begin{center}
  \includegraphics[width=8cm,height=6cm]{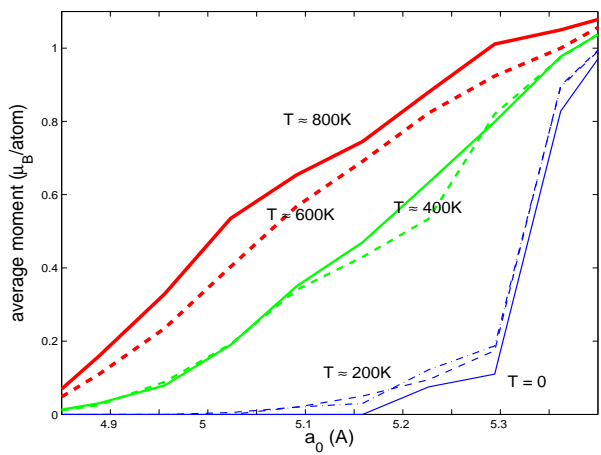}
  \end{center}
  \caption{The average magnetic moment as function of lattice constant, $a_0$, for different levels of
  disorder ($\sigma = u/a_0$) between 0 ("ordered structure") and 0.063.} 
  \label{figcem}
  \end{figure}

Calculations for the ordered cell, and for the 1-atom fcc cell, show that a ferromagnetic (FM) ground state develops
when the lattice constant $a_0$ exceeds about 5.2 \AA~ \cite{ce}. The ground state solutions shift
easily between a low-magnetic ($m \approx 0$) and a high-magnetic state ($m$ $\geq 0.4 \mu_B/atom$)
when $a_0 \sim 5.3$ \AA.
The state at the absolute minimum of the total energy $E_0$ is non-magnetic (NM), near $a_0 = 4.79 \AA$,
compared to 4.62 \AA~ when using LSDA \cite{ce}.
The experimental values at RT are near 4.85~\AA~ and 5.16~\AA~ for the $\alpha$- and $\gamma$-phase,
respectively \cite{fran}.
From the Stoner model it is expected that FM develops at large volume, when the band narrowing makes  
$N(E_F)$ larger. The gain in exchange energy overcomes the loss of kinetic energy at the FM transition \cite{c15},
but the Coulomb interaction can make small corrections to this energy balance \cite{ce98}.
These effects are included in the self-consistent, spin-polarized calculations. 
Structural disorder introduces local volume fluctuations, and the degree of localization of f-electrons
depends directly on the surroundings. In Fig. \ref{fig1} is shown an example of
the correlation between local volume variations ($<d_{nn}>$, which is defined as
the average of the 12 nearest-neigbor distances
around each of the 32 sites), and the site decomposed $N(E_F)$-values, as well as the local moments,
for a case with $\sigma=0.04 a_0$, $a_0$=5.29 \AA. As seen, when $<d_{nn}>$ is considerably larger ($>$ 0.72)
than the value for ordered lattice (0.7071), $N(E_F)$ and $m$ are highest. 
The valence charge per Ce varies quite linearly from 10.4 el./Ce for the sites with the lowest moment 
to about 9.7 el./Ce when the moment is just above 1 $\mu_B$/Ce.
Disorder has a supplementary effect
on the average moment if the lattice constant
is below the critical value for  a high moment: Since the local volume and the moments are increased
on many of the sites with large $<d_{nn}>$, it leads to an increase of the total FM moment of the cell. That some of the
sites are "compressed" (small $<d_{nn}>$) does not reduce the total moment, because their local moments are small already
in the ordered lattice.
Oppositely, at large $a_0$ when the moment for the ordered structure is close to its maximum, about 1.1 $\mu_B/atom$,
there is no (or very little) effect on the total moment from disorder. The saturation of the total spin
moment near 1  $\mu_B/atom$ for one occupied f-electron can be understood from Hund's first rule. Thus, disorder fluctuations
can not make the local moment much higher even if the local volume is increased, but local compressions
could rather decrease the moment. This explains the saturation of the highest local moments seen in fig \ref{fig1}
for the sites with the highest $<d_{nn}>$ and $N(E_F)$.
Fig. \ref{figcem} displays the averaged $m(a_0)$ for different distortion amplitudes. As seen, near the transition region there
are large effects on magnetism because of lattice disorder. Magnetism appears suddenly at
$a_0 \geq 5.3$~\AA~ for the ordered structure, but is much more gradual and starts at lower volume 
when the disorder is large.
Therfore,  vibrational disorder is crucial 
for a good understanding of the properties of Ce. 

All self-consistent calculations start from the potential for the ordered structure. The final
FM configurations converge gradually during the iterations. In a few cases, usually when the magnetic
moments are small, it is possible to find sites where the spin orientation is opposite to the majority moments,
as if anti-ferromagnetism (AFM) was installed locally. These solutions develop very slowly, but they seem not to 
concern large-moment cases at large volume and large temperature. Therefore, such solutions 
are not important for the free-energy arguments in the next section. The tendencies
for local AFM diminish when the electronic temperature is raised. Thus, smearing due to the
Fermi-Dirac distribution is not favorable to spin flips, while details in the local environment caused by
lattice disorder can be so.

\subsection{Free energies}
  
Three sources of entropy were included in the GGA calculations 
for ordered fcc Ce \cite{ce}; vibrational, electronic and magnetic. 
The difference in vibrational free energy at two volumes $V_i$ is
\begin{equation}
\Delta F_{vib}(V) = ~ 3 k_B T ~  ln(\frac{\Theta_{\gamma}}{\Theta_{\alpha}})
\end{equation}
where the Debye temperatures
$\Theta_i$ are closely related to $\sqrt{(V_i^{1/3} B_i)}$, where $B_i$ are
the Bulk moduli of the two phases. The latter are calculated to be in the range 15-20 GPa for FM
Ce and 20-30 GPa for NM Ce. This agrees with experiment \cite{lip}, but it is delicate to determine
the full $T,P$-dependence from $\it{ab-initio}$ calculations because of the sharp drop
of $B_i$ at the transition. Here we choose to fit $\Theta_D$ to the experimental
results in ref. \cite{lip}. This gives 
$\Theta_D \approx 160- 22 \cdot (1+sign(m-\frac{1}{2}) sin(\pi(m-\frac{1}{2}))^{\frac{1}{4}}) + 15 \cdot (3.54 -a_0)$
 (in $K$). The last term makes the continuous decrease of $\Theta_D$ from about 160$K$ at small volume to
 about 140K at large volume. This is the typical behavior in almost all materials, since $B$ normally decreases with
 increasing volume. The second term is introduced in order to include a sharp discontinuity ($\sim30K$) in
 $\Theta_D$ at the transition when $m$ is close to 0.5 $\mu_B$/atom. Thus,
vibrations favor 
the $\gamma$-phase because of its softer lattice. 

 \begin{figure}[h]
  \begin{center}
  \includegraphics[width=8cm,height=6cm]{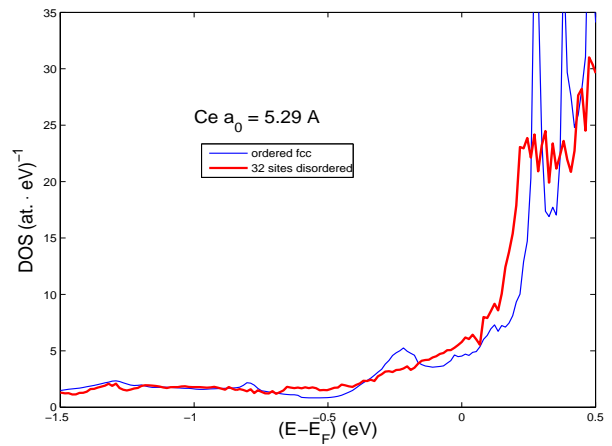}
  \end{center}
  \caption{The DOS near $E_F$ in the ordered and disordered 32 cells at the lattice
  constant $a_0 =$ 5.29 \AA. The disordered case has $\sigma$=0.04.} 
  \label{fig3}
  \end{figure}

The electronic entropy is calculated as:
\begin{equation}
S_{el}= -\int N(E) (f ln f + (1-f) ln(1-f)) dE
\end{equation}
where $N(E)$ is the electronic density-of-states and $f$ the Fermi-Dirac distribution. This quantity
is almost proportional to $N(E_F)$ ($S_{el} \approx \frac{2\pi^2}{3} k_B^2T N(E_F)$), and
as shown in Fig. \ref{fig3}, disorder makes 
$N(E_F)$ larger. The calculated $F_{el} = E_{tot}-TS_{el}$
 favors the $\gamma$ phase
even more than what was found in ref \cite{ce}, because of disorder.

A large entropy comes from fluctuations of magnetic moments,
\begin{equation}
S_{m}= k_B ~ ln [2 (L-m/2) +1] 
\end{equation}
which includes an orbital moment $L$, and a spin part being half of the magnetic moment, $m/2$.
A full moment of a single f-electron makes $L$=3 according to Hund's first rule.
Here we apply atomic-like Paschen-Back calculations, which for the spin-orbit coupling in Ce-f give
$L \approx 2.5 \cdot m$ for $m \leq $1 \cite{ce}.
The moments $L$ and $m$ make
a substantial entropy contribution at large $T$, which stabilizes the $\gamma$-phase depending
on the evolution of $m(T,V)$.
Without consideration of disorder  $m(T,V)$ follows the thin full line in Fig. \ref{figcem},
which is the basis for the result in ref. \cite{ce}. As seen in Fig. \ref{figcem}, disorder moves
the magnetic transition towards lower volume. This fact makes the magnetic entropy contribution larger, and
disorder is therefore important for the $\alpha-\gamma$
transition. 
Entropy from phase mixing \cite{sva} is not accounted for in the present work.

 \begin{figure}[h]
  \begin{center}
  \includegraphics[width=8cm,height=6cm]{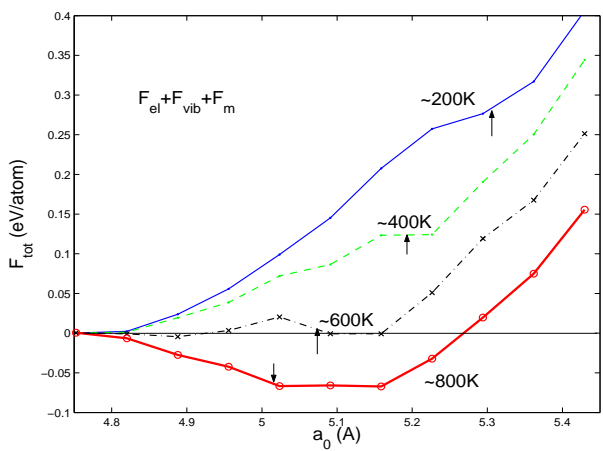}
  \end{center}
  \caption{Total relative free energies ($F_{tot}$ in $eV/atom$) as function of 
  the lattice constant, $a_0$. The (blue) thin full line
  shows the total free energy at low $T$, when the disorder is dominated by ZPM. The subsequent lines
  show the results at 400K, 600K and 800K, respectively. The small vertical arrows indicate the
  volumes where the average magnetic moment exceeds 0.5 $\mu_B$ per atom.} 
  \label{figet}
  \end{figure}
  
The electronic ($F_{el}$) free energy is calculated from the 1 atom/cell results with T-dependences
entering through the Fermi-Dirac distribution and a DOS broadening.
The two phases do not co-exist at equal volumes. The minimum of $E_{tot}$ at 4.79 ~\AA~ is for
NM $\alpha$-Ce.
FM grows gradually as $a_0 \gtrsim$ 5.2 ~\AA, but there is no second local minimum of $E_{tot}$
even if there is some lowering of $E_{tot}$ beyond 5.3 ~\AA, see Fig \ref{figet}.
The signature of FM, seen the electronic free energy curve, 
moves to lower volume because of higher moments when $T$ increases. Disorder at $T \sim 400K$ makes moments to 
appear already below $a_0=$  5 \AA~.
 
The next step is to add the vibrational entropy and magnetic entropies 
from the disordered large cell calculations (scaled by 1/32) to the electronic free energies from
the 1 atom/cell results. The results for $T$ between 200 and 800 K are shown
in  Fig \ref{figet}. The crossover from the NM low volume phase to the FM fluctuating phase at large
volume ($a_0$ in the range 5.1-5.15~\AA) occurs just below 600 K. These results are
in closer agreement with experiment than in the calculations without
consideration of disorder \cite{ce}. 
The electronic total energy goes down with increasing moment,
and the onset of magnetism (indicated by the arrows in Fig \ref{figet}) is seen to coincide with a small
discontinuity in the $F_{tot}$-curves, which moves towards lower volume as $T$ is increasing. 
Since the average moments in the disordered
(supercell) results
are higher than the moments in ordered Ce, it can be expected that
the $\gamma$-phase will be stabilized further.

As noted earlier, the equilibrium volume for the NM ground state is better from GGA than from LSDA.
Since the crossover to the high volume FM state occurs at a reasonable T, when using free energies
from GGA, gives us confidence that GGA is reliable also in the FM regime. Results using GGA+U (and LSDA+U)
have the total energies for the FM state lower than for the NM state already at $T=0$ when U is large, 
which is incorrect \cite{tran}.
Other properties such as moments and DOS at $E_F$ seem to depend less on the exact choice of U \cite{tran},
so the free energy contribution at large $T$ should be comparable to the present results. Thus, consideration
of disorder and entropies in addition to GGA+U could easily destroy the good $T$-dependence if correlation
is imposed by having U larger than $\sim$1 eV.

\subsection{Core levels}

Core level energies are probes of potential shifts and can be used to measure
effects of disorder and magnetic fluctuations. The local variations of exchange splitting (proportional
to the local magnetic moments)
and Madelung shifts caused by disorder show up as broadening of spectroscopic ensembles of core levels.
 This opens a possibility for core level spectroscopy to identify the effects of disorder. 
 For instance, in the NM high pressure $\alpha$-phase at RT, disorder is calculated to make a broadening
 of the 4s level  of about 0.16 eV. 
By removing the pressure (but keeping $T$ constant) to get the magnetic $\gamma$-phase, these
 levels broaden to about 0.42 eV because of the variations of the local moments (the averaged moment is 0.36
$\mu_B/atom$). 
Without disorder there would be no variations of the Madelung shifts, and identical  exchange splittings on
 all sites should produce two sharp lines separated by 0.16 eV for a moment of 0.36 $\mu_B/atom$. The 
broadening from disorder is too large for a clear identification of separated spin up and down peaks.
These broadenings do not include other smearing mechanisms due to the experimental method or
other types of lattice imperfections.
 The Madelung shifts and exchange splitting of the 4p and 4d levels are comparable, with spin-orbit splittings
 of 18.9 and 3.3 eV, respectively.

\section{Conclusion}

All entropies contribute to a crossover from the $\alpha$- to the $\gamma$-phase at about 800 K 
when using GGA without effects from disorder \cite{ce}.  Here,
with disorder, the transition is calculated to occur below 600K,
at a volume in better agreement with experiment. 
Entropies and the effects of disorder exist always, and they should be considered even in calculations based 
on strong correlation. The behavior at $T=0$
is not certain, especially because of the dhcp $\beta$-phase that 
replaces the $\gamma$-phase at low $T$ on the $P=0$ line \cite{fran}. Nevertheless, 
ignoring this and doing an extrapolation towards low $T$ on the $\alpha-\gamma$ separation line of
the phase diagram suggests a negative transition pressure at $T=0$. This can be taken as a support 
for potentials with large correlation 
 \cite{casa,lana,tran}, but it is not clear how of such results will behave at high $T$.  
Most observations of the $\alpha-\gamma$
transition are made in the range 150-450 K, and it is important to test theoretical results 
in the similar temperature range.
The fact that the transition can be described quite accurately
by temperature dependent DFT calculations with thermal disorder and entropies, 
is a strong support for standard DFT. 
Note that DFT has been applied successfully to a vast amount of metallic systems without
relying on adjustable parameters for correlation. 
Finally, Ce can be added to the list of materials where thermal disorder 
is seen to be important for the physical properties.

  \end{document}